\documentclass[conference,a4paper]{IEEEtran}
\usepackage{xcolor}
\usepackage{balance}
\usepackage[nolist]{acronym}

\ifCLASSINFOpdf
  \usepackage[pdftex]{graphicx}
\else
  \usepackage[dvips]{graphicx}
\fi
\ifCLASSOPTIONcompsoc
 \usepackage[caption=false,font=normalsize,labelfont=sf,textfont=sf]{subfig}
\else
 \usepackage[caption=false,font=footnotesize]{subfig}
\fi
\begin{document}

%Para evitar los cortes de palabra
\sloppy
\begin{acronym}

\acro{2D}{Two Dimensions}%
\acro{2G}{Second Generation}%
\acro{3D}{Three Dimensions}%
\acro{3G}{Third Generation}%
\acro{3GPP}{Third Generation Partnership Project}%
\acro{3GPP2}{Third Generation Partnership Project 2}%
\acro{4G}{Fourth Generation}%
\acro{5G}{Fifth Generation}%

%%%%%A%%%%%
\acro{AI}{Artificial Intelligence}%
\acro{AoA}{Angle of Arrival}%
\acro{AoD}{Angle of Departure}%
\acro{AR}{Augmented Reality}%
\acro{AP}{Access Point}
\acro{AE}{Antenna Element}
%%%%%B%%%%%

\acro{BER}{Bit Error Rate}%
\acro{BPSK}{Binary Phase-Shift Keying}%
\acro{BRDF}{ Bidirectional Reflectance Distribution Function}%
\acro{BS}{Base Station}%

%%%%%C%%%%%
\acro{CA}{Carrier Aggregation}%
\acro{CDF}{Cumulative Distribution Function}%
\acro{CDM}{Code Division Multiplexing}%
\acro{CDMA}{Code Division Multiple Access}%
\acro{CI}{close-in free space reference distance}
\acro{CPU} {Central Processing Unit}
\acro{CUDA}{Compute Unified Device Architecture}
\acro{CDF}{Cumulative Distribution Function}
 
%%%%%D%%%%%
\acro{D2D}{Device-to-Device}%
\acro{DL}{Down Link}%
\acro{DS}{Delay Spread}%
\acro{DAS}{Distributed Antenna System}
\acro{DKED}{double knife-edge diffraction}
\acro{DUT}{Device Under Test}

%%%%%E%%%%%

\acro{EDGE}{Enhanced Data rates for GSM Evolution}%
\acro{EIRP}{Equivalent Isotropic Radiated Power}%
\acro{eMBB}{Enhanced Mobile Broadband}%
\acro{eNodeB}{evolved Node B}%
\acro{ETSI}{European Telecommunications Standards Institute}%
\acro{ER}{Effective Roughness}%
\acro{E-UTRA}{Evolved UMTS Terrestrial Radio Access}%
\acro{E-UTRAN}{Evolved UMTS Terrestrial Radio Access Network}%
\acro{EF}{Electric Field}

%%%%%F%%%%%
\acro{FDD}{Frequency Division Duplexing}%
\acro{FDM}{Frequency Division Multiplexing}%
\acro{FDMA}{Frequency Division Multiple Access}%
\acro{FoM}{Figures of Merit}

%%%%G%%%%%
\acro{GI}{Global Illumination} %
\acro{GIS}{Geographic Information System}%
\acro{GO}{Geometrical Optics} %
\acro{GPU}{Graphics Processing Unit}%
\acro{GPGPU}{General Purpose Graphics Processing Unit}%
\acro{GPRS}{General Packet Radio Service}%
\acro{GSM}{Global System for Mobile Communication}%

%%%%%H%%%%%
\acro{H2D}{Human-to-Device}%
\acro{H2H}{Human-to-Human}%
\acro{HDRP}{High Definition Render Pipeline}
\acro{HSDPA}{High Speed Downlink Packet Access}
\acro{HSPA}{High Speed Packet Access}%
\acro{HSPA+}{High Speed Packet Access Evolution}%
\acro{HSUPA}{High Speed Uplink Packet Access}
\acro{HPBW}{Half-Power Beamwidth}

%%%%%I%%%%%
\acro{IEEE}{Institute of Electrical and Electronic Engineers}%
\acro{InH}{Indoor Hotspot} %
\acro{IMT} {International Mobile Telecommunications}%
\acro{IMT-2000}{\ac{IMT} 2000}%
\acro{IMT-2020}{\ac{IMT} 2020}%
\acro{IMT-Advanced}{\ac{IMT} Advanced}%
\acro{IoT}{Internet of Things}%
\acro{IP}{Internet Protocol}%
\acro{ITU}{International Telecommunications Union}%
\acro{ITU-R}{\ac{ITU} Radiocommunications Sector}%
\acro{IS-95}{Interim Standard 95}%
\acro{IES}{Inter-Element Spacing}
\acro{IA}{Interferer Array}

%%%%%J%%%%%

%%%%%K%%%%%
\acro{KPI}{Key Performance Indicator}%

%%%%%L%%%%%
\acro{LB} {Light Bounce}
\acro{LIM}{Light Intensity Model}%
\acro{LoS}{line of sight}%
\acro{LTE}{Long Term Evolution}%
\acro{LTE-Advanced}{\ac{LTE} Advanced}%
\acro{LSCP}{Lean System Control Plane}%
\acro{LSI} {Light Source Intensity}

%%%%%M%%%%%
\acro{M2M}{Machine-to-Machine}%
\acro{MatSIM}{Multi Agent Transport Simulation}
\acro{METIS}{Mobile and wireless communications Enablers for Twenty-twenty Information Society}%
\acro{METIS-II}{Mobile and wireless communications Enablers for Twenty-twenty Information Society II}%
\acro{MIMO}{Mul\-ti\-ple-In\-put Mul\-ti\-ple-Out\-put}
\acro{mMIMO}{massive MIMO}%
\acro{mMTC}{massive Machine Type Communications}%
\acro{mmW}{millimeter-wave}%
\acro{MU-MIMO}{Multi-User MIMO}
\acro{MMF}{Max-Min Fairness}
\acro{MKED}{Multiple Knife-Edge Diffraction}
\acro{MF}{Matched Filter}
\acro{mmWave}{Millimeter Wave}
\acro{MA}{Main Array}

%%%%%N%%%%%
\acro{NFV}{Network Functions Virtualization}%
\acro{NLoS}{non line of sight}%
\acro{NR}{New Radio}%
\acro{NRT}{Non Real Time}%
\acro{NYU}{New York University}%

%%%%%O%%%%%
\acro{O2I}{Outdoor to Indoor}%
\acro{O2O}{Outdoor to Outdoor}%
\acro{OFDM}{Orthogonal Frequency Division Multiplexing}%
\acro{OFDMA}{Or\-tho\-go\-nal Fre\-quen\-cy Di\-vi\-sion Mul\-ti\-ple Access}
\acro{OtoI}{Outdoor to Indoor}%
\acro{OTA}{Over-The-Air}

%%%%%P%%%%%
\acro{PDF}{Probability Distribution Function}
\acro{PDP}{Power Delay Profile}
\acro{PHY}{Physical}%
\acro{PLE}{Path Loss Exponent}

%%%%%Q%%%%%
\acro{QAM}{Quadrature Amplitude Modulation}%
\acro{QoS}{Quality of Service}%

%%%%%R%%%%%
\acro{RCSP}{Receive Signal Code Power}
\acro{RAN}{Radio Access Network}%
\acro{RAT}{Radio Access Technology}%

%\acro{RB}{Radio Bearer}%
\acro{RAN}{Radio Access Network}%
\acro{RMa}{Rural Macro-cell}%
\acro{RMSE} {Root Mean Square Error}
\acro{RSCP}{Receive Signal Code Power}%
%\acro{RT}{Real Time}%
\acro{RT}{Ray Tracing}
\acro{RX}{receiver}
\acro{RMS}{Root Mean Square}
\acro{Random-LOS}{Random Line-Of-Sight}
\acro{RF}{Radio Frequency}

%%%%%S%%%%%
\acro{SB} {Shadow Bias}
\acro{SC}{small cell}
\acro{SDN}{Software-Defined Networking}%
\acro{SGE}{Serious Game Engineering}%
\acro{SF}{Shadow Fading}%
\acro{SIMO}{Single Input Multiple Output}%
\acro{SINR}{Signal to Interference plus Noise Ratio}
\acro{SISO}{Single Input Single Output}%
\acro{SMa}{Suburban Macro-cell}%
\acro{SNR}{Signal to Noise Ratio}
\acro{SU}{Single User}%
\acro{SUMO}{Simulation of Urban Mobility}
\acro{SS} {Shadow Strength}

%%%%%T%%%%%

\acro{TDD}{Time Division Duplexing}%
\acro{TDM}{Time Division Multiplexing}%
\acro{TD-CDMA}{Time Division Code Division Multiple Access}%
\acro{TDMA}{Time Division Multiple Access}%
\acro{TX}{transmitter}
\acro{TZ}{Test Zone}

%%%%%U%%%%%

\acro{UAV}{Unmanned Aerial Vehicle}%
\acro{UE}{User Equipment}%
\acro{UI}{User Interface}
\acro{UHD}{Ultra High Definition}
\acro{UL}{Uplink}%
\acro{UMa}{Urban Macro-cell}%
\acro{UMi}{Urban Micro-cell}%
\acro{uMTC}{ultra-reliable Machine Type Communications}%
\acro{UMTS}{Universal Mobile Telecommunications System}%
\acro{UPM}{Unity Package Manager}
\acro{UTD}{Uniform Theory of Diffraction} %
\acro{UTRA}{{UMTS} Terrestrial Radio Access}%
\acro{UTRAN}{{UMTS} Terrestrial Radio Access Network}%
\acro{URLLC}{Ultra-Reliable and Low Latency Communications}%

%%%%%V%%%%%
\acro{V2V}{Vehicle-to-Vehicle}%
\acro{V2X}{Vehicle-to-Everything}%
\acro{VP}{Visualization Platform}%
\acro{VR}{Virtual Reality}%
\acro{VNA}{vector network analyzer}
\acro{VIL}{Vehicle-in-the-loop}

%%%%%W%%%%%
\acro{WCDMA}{Wideband Code Division Multiple Access}%
\acro{WINNER}{Wireless World Initiative New Radio}%
\acro{WINNER+}{Wireless World Initiative New Radio +}%
\acro{WiMAX}{Worldwide Interoperability for Microwave Access}%
\acro{WRC}{World Radiocommunication Conference}%

%%%%%X%%%%%
\acro{xMBB}{extreme Mobile Broadband}%

%%%%%Z%%%%%
\acro{ZF}{Zero Forcing}

\end{acronym}

\title{Impact of Excitation and Weighting Errors on Performance of Compact OTA Testing Systems}

\author{\IEEEauthorblockN{
Alejandro Antón Ruiz\IEEEauthorrefmark{1} and   % 1st author, 1st affiliations
Andrés Alayón Glazunov\IEEEauthorrefmark{1}   % 2nd author, 2nd affiliations
}                                     % ...
%\\
\IEEEauthorblockA{\IEEEauthorrefmark{1}% 1st affiliations
Department of Electrical Engineering, University of Twente, Enschede, The Netherlands}
}



\maketitle

\begin{abstract}

This paper investigates the impact of complex excitation errors of the chamber array antenna on the accuracy of the test zone of a random line-of-sight over-the-air testing setup. First, several combinations of compact chamber arrays of lengths $L$ and short distances $D$ between the test zone and the chamber array, which emulate a plane wave impinging at the test zone are obtained. The chamber array is linear and uniform with 100 antenna elements, and a linear taper was applied to some of the elements to emulate a plane wave impinging at the test zone with more compact setups. A subset of $L$ and $D$ was chosen, providing compact over-the-air test setups that fulfilled the defined figures of merit, which assess the similarity of the obtained field distribution to that of a plane wave. The tolerance of the chosen setups to complex excitation errors of the chamber array was then investigated, concluding that these errors must be considered when defining appropriate $L$ and $D$ combinations. Moreover, the performance of the matched filter and zero-forcing algorithms is evaluated for errors of the device under test array weighting coefficients. A random line-of-sight over-the-air testing setup with two arrays was simulated, where one of the arrays emulated the desired signal and the other emulated the interference, observing that the errors were more significant at higher signal-to-noise ratios. Additionally, the zero-forcing algorithm was more sensitive to errors than the matched filter, which was expected since the accuracy of the former for interference suppression is critical.

\end{abstract}

\vskip0.5\baselineskip
\begin{IEEEkeywords}
 OTA, automotive, precoding, excitation errors.
\end{IEEEkeywords}

% \fi
% \IEEEpeerreviewmaketitle

\section{Introduction}

\ac{OTA} testing has become the standard method for full performance evaluation of wireless devices. It accounts for the antenna characteristics of the \ac{DUT} in an environment that emulates its actual use. Besides testing the communication protocols and the performance of the radio frequency part, it may also consider other sources of error such as using its own power source \cite{OTA_adv}.

\ac{OTA} is a key enabler of the development of the automotive industry, especially as it is moving towards the integration of an increasing number of sensors, i.e., radars, lidars, cameras, as well as wireless communications and GPS. Radars will be mostly used in the $76-81$~GHz range since the $24$~GHz ultra-wideband has been phased out this January \cite{ETSI_24_phase_out}. However, many other products still operate in the lower \ac{mmWave} bands and below, including \ac{V2X} communications, which operate in the sub-6 GHz bands for now. Nevertheless, there is a need for larger data rates than those achieved by sub-6 GHz bands to support applications such as exchanging raw data from sensors in vehicles. This can be achieved by resorting to the \ac{mmWave} frequencies, such as the already defined FR2 frequency bands \cite{Justif_mmWave}.

Currently, there are \ac{OTA} testing systems available for \ac{mmWave} communications, e.g., \cite{Bluetest} and \cite{RohdeSchwarz_Comm}. There are already solutions for \ac{mmWave} radar testing too \cite{RohdeSchwarz_Radar}, including car-mounted ones \cite{VISTA_ILMENAU_RADAR}. However, to the best of the authors' knowledge, and in agreement with \cite{OTA_adv}, there are no available solutions for FR2 communications automotive \ac{OTA} testing that are feasible in terms of hardware costs and dimensions, so further efforts must be made to devise such solutions.

In this paper we present numerical simulations of an \ac{OTA} system at $28$~GHz, corresponding to the center frequency of the 3GPP n257 band, chosen as a representative of FR2. One of the main challenges for automotive \ac{OTA}, especially for vehicle-in-the-loop testing, is fulfilling the far-field criterion, i.e., the Fraunhofer distance. For a whole car at \ac{mmWave}, it extends to several km. Clearly, these distances are not feasible in a controlled environment. Thus, there is a need to resort to \ac{OTA} techniques that, while not physically being in the far-field, can emulate an impinging plane wave at the automobile. Among such techniques, there are compact test ranges, plane wave generators and random line-of-sight (usually denoted in the literature as RLOS, Random-LOS, or Ranlos). We take the approach of the Random-LOS technique \cite{MSK2,MSK3,MSK1}.

In this work, we first conduct a study of the combined effects of linear chamber array size and the distance between the center of the chamber array and the center of the \ac{TZ} for a given \ac{TZ} size. The performance criteria is a set of \ac{FoM} that evaluate the similarity of the field emulated in the \ac{TZ} to a plane wave. The idea is to find the most compact set, i.e., the smallest array and shortest distance (at least shorter than Fraunhofer distance) that satisfies the accuracy criteria. We also study the tolerance in terms of \ac{FoM} compliance of a subset of distances and chamber array sizes to random complex excitation errors of the chamber array due to, e.g., manufacturing tolerances and quantization errors. We also study the impact of errors on the performance of \ac{MF} and \ac{ZF} algorithms due to complex errors in the weights of the \ac{DUT} array. The results show that chamber array excitation errors must be considered when selecting distances and chamber array sizes, and also that \ac{DUT} errors affect significantly more to \ac{ZF} than to \ac{MF}.

\section{\ac{OTA} setup and \ac{FoM} for \ac{TZ} quality evaluation}
\label{Section II}

\subsection{OTA setup}
\label{Section II A}
\begin{figure}
\centering
\includegraphics[width=0.5\columnwidth]{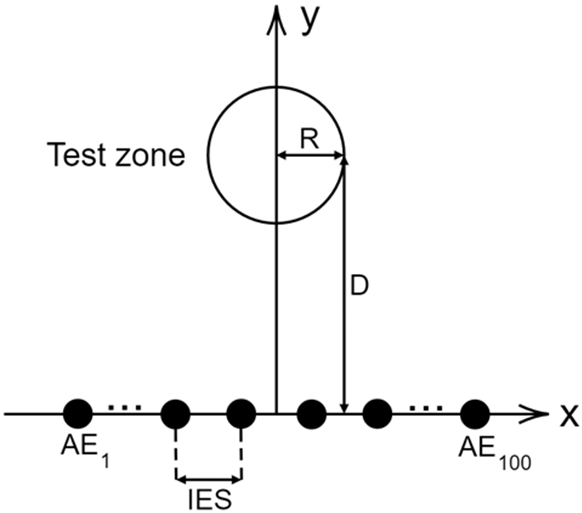}
\caption{Initial OTA setup}
\label{F1}
\end{figure}

The basic arrangement of the \ac{OTA} setup is depicted in Fig.~\ref{F1}, which is not to scale. The chamber array is defined as a uniform linear array along the $x-$axis, with a fixed number of \acp{AE} $N_C=100$. The \acp{AE} are idealized vertically polarized isotropic radiators operating at $28$~GHz. A linear taper from 0 dB to -6 dB is applied to 25 elements on each side of the array to reduce field fluctuations in the \ac{TZ} as explained in Section \ref{Section II C} \cite{P_ANDRES_TAPER}. Thus, the \ac{EF} at a given point $P$, accounting only for the vertical polarization ($z-$axis), is computed by the superposition principle as
\begin{equation}
\label{Eq1}
    E_z=\sum_{i=1}^{N_C}{t_{c_i}E_0\frac{e^{-jkr_i}}{4\pi r_i}}=\sum_{i=1}^{N_C}{t_{c_i}\frac{e^{-jkr_i}}{4\pi r_i}},
\end{equation}
where $E_0$ is set to $1$ for simplicity, $t_{c_i}={10}^{t_{c_{i_{dB}}}/20}$ is the tapering coefficient in linear scale of the $i$-th \ac{AE}, being $t_{c_{i_{dB}}}$ the dB-scale tapering coefficient, and $r_i$ is the distance between the $i$-th \ac{AE} and $P$. \ac{IES} is considered a variable, ranging from $0.5\lambda$ to $1.5\lambda$, with $0.05\lambda$ step, resulting in a variable chamber array length $L=(N_C-1)IES$. The \ac{TZ} is contained by the $XY-$plane and it is defined as a circle of radius $R=(N_C-1)IES/4$, where \ac{IES} equals $0.5\lambda$. Hence, $R=99\lambda/8=13.26$ cm is a quarter of the length of the shortest considered chamber array. The center of the \ac{TZ} is at a distance $D$ of the chamber array, along the $y-$axis.

\subsection{FoM for the \ac{TZ}}
\label{Section II B}

The objective of this \ac{OTA} setup is to emulate an \ac{EF} distribution over the \ac{TZ} that emulates a plane wave. To assess the accuracy of the plane wave, several \ac{FoM} are defined. First, we consider
\begin{equation}
    R_{mag}=max{\left(20\log_{10}{\left(\left|\mathbf{E}_z\right|\right)} \right)}-min{\left(20\log_{10}{\left(\left|\mathbf{E}_z\right|\right)}\right)},
\end{equation}
where $\mathbf{E}_z$ is the \ac{EF} of every sample belonging to the \ac{TZ}. $R_{mag}$ defines the dynamic range of the magnitude of the \ac{EF} samples in the \ac{TZ}. These samples come from the nodes of a mesh with equal $\lambda/8$ interval in both the $x-$ and $y-$axes, so the density of samples is constant over the circular \ac{TZ} area.

Secondly, we evaluate the standard deviation of the magnitude of the \ac{EF} samples in the \ac{TZ} in dB
\begin{equation}
    \sigma_{mag}=\sqrt{\frac{\sum_{s=1}^{Ns}\left(X_s-\bar{x}\right)^2}{N-1}},
\end{equation}
where $N$ is the number of samples in the \ac{TZ}, $X_s$ is the \ac{EF} magnitude in logarithmic units of the $s$-th sample, and $\bar{x}$ is the mean of the \ac{EF} magnitude over all the \ac{TZ} samples. This \ac{FoM} is supported by the 3GPP \cite{REF_sigma_mag_value}.

Thirdly, we compute the dynamic range of the phase of the \ac{EF} over the \ac{TZ}
\begin{equation}
    R_{phs_{rows_n}}=\max{\left(\angle{\mathbf{E}_{z_n}}\right)}-\min{\left(\angle{{\mathbf{E}_z}_n}\right)},
\end{equation}
\begin{equation}
    R_{phs}=\max{\left({\mathbf{R}_{phs}}_{rows}\right)},
\end{equation}
where $\mathbf{R}_{phs}$ contains the phase range over each parallel stripe (with respect to the chamber array) of samples of the \ac{TZ}, $R_{phs_{rows_n}}$ is the phase range of a given parallel stripe, $\mathbf{E}_{z_n}$ contains the \ac{EF} values of a given parallel stripe, and $\angle$ denotes the phase value or angle. This \ac{FoM} may arise a concern due to the periodic nature of the phase. Indeed, if wrapped up to the interval $[0^{\circ},360^{\circ}[$, then one could argue that, e.g. if within a row, there is a phase value of 359$^{\circ}$ and other value of 2$^{\circ}$, the resulting $R_{phs_{rows_n}}$ would be 357$^{\circ}$. This has been taken into account, so that the correct variation, of 3$^{\circ}$ in this case, is always computed. We limit the variation to 180$^{\circ}$ because that is the maximum phase deviation that can actually occur.

It is worth noting the ideal values of the \ac{FoM}. Since the desired \ac{EF} distribution is that of a plane wave, the magnitude of the \ac{EF} over the \ac{TZ} should be the same, so $R_{mag}$ and $\sigma_{mag}$ should be 0 dB. Similarly, the phase should be constant along each parallel stripe, so $R_{phs}=0^{\circ}$. However, a perfect plane wave \ac{EF} distribution is not achievable due to physical limitations, e.g., finite aperture and a number of sources. Thus, we focus on acceptable \ac{FoM} values: $R_{mag} \leq 1$~dB is commonly accepted \cite{REF_R_mag_value}, while $\sigma_{mag} \leq 0.25$~dB is, according to \cite{REF_sigma_mag_value}, required, and $R_{phs} \leq 10^{\circ}$ is often assumed as an acceptable limit \cite{Hybrid_exci}.

\subsection{L and $D$ satisfying the \ac{FoM} limits}
\label{Section II C}

\begin{figure}
    \centering
        \subfloat[]{\includegraphics[width=0.7\columnwidth]{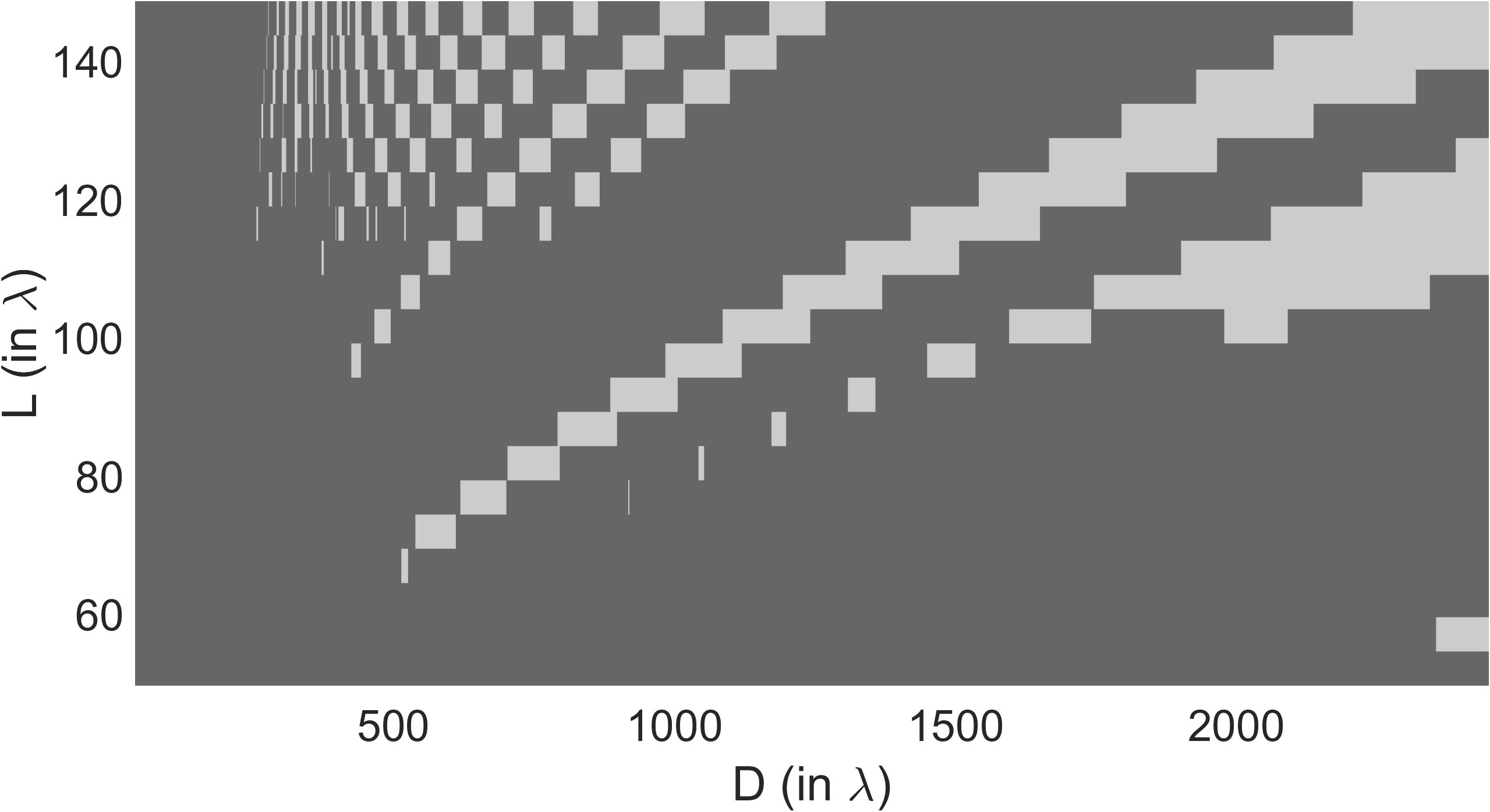}}
        \label{F2a}
        \subfloat[]{\includegraphics[width=0.7\columnwidth]{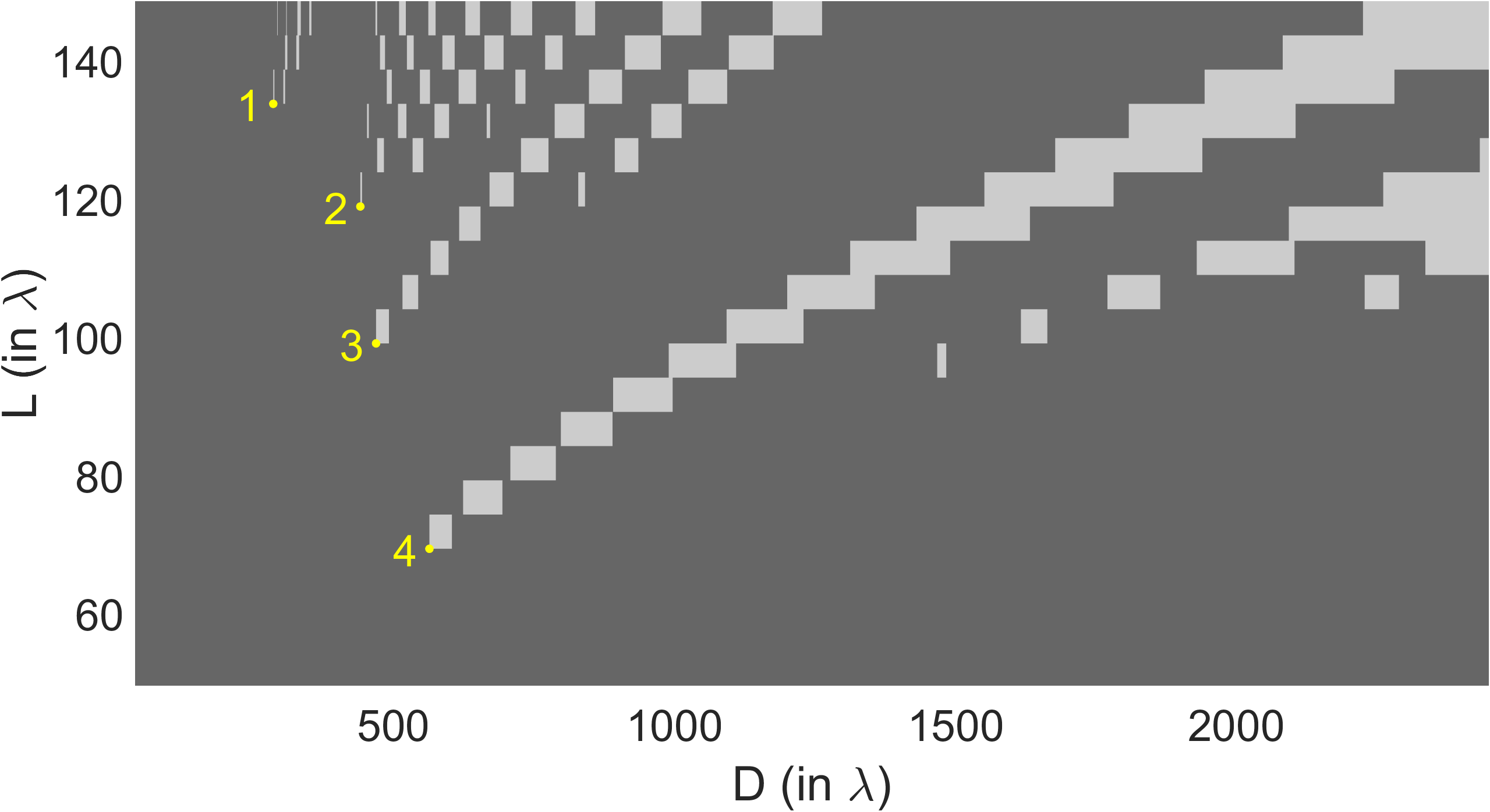}}
        \label{F2b}
        \subfloat[]{\includegraphics[width=0.7\columnwidth]{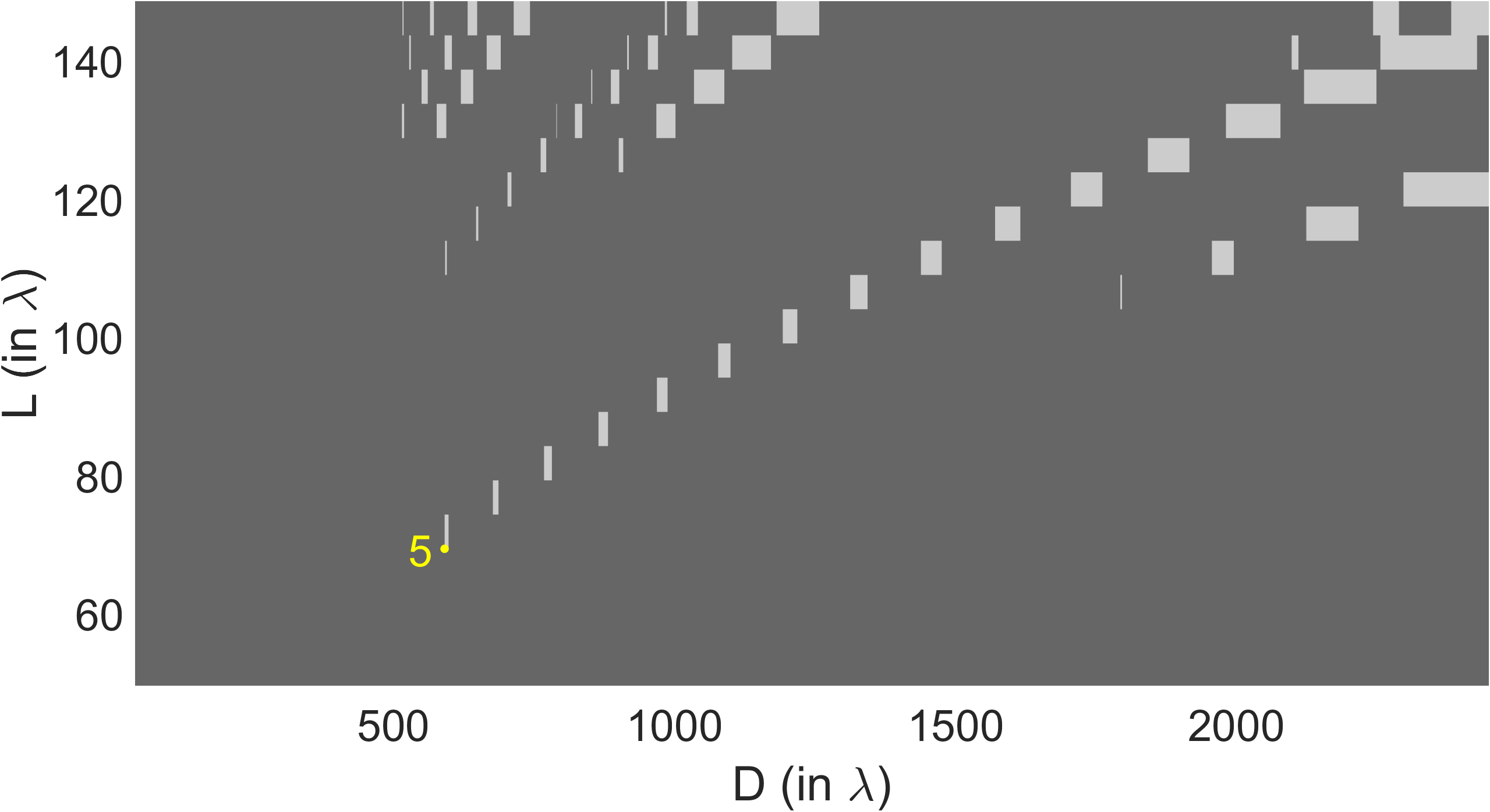}}
        \label{F2c}
    \caption{$L$ and $D$ values compliant with all \ac{FoM} limits in light grey. (a) $\sigma_{mag}=0.25$ dB, $R_{mag}=1$ dB, $R_{phs}=10^{\circ}$, (b) $\sigma_{mag}=0.225$ dB, $R_{mag}=0.9$ dB, $R_{phs}=9^{\circ}$, (c) $\sigma_{mag}=0.2$ dB, $R_{mag}=0.8$ dB, $R_{phs}=8^{\circ}$. Note that in (b) and (c), the points used for Sections \ref{Section III} and \ref{Section IV} are marked.}
    \label{F2}
\end{figure}

Having defined the \ac{FoM} and their acceptable values, we investigate the $L$ and $D$ combinations that fulfill them for the considered \ac{OTA} setup. This study is extended to stricter acceptable values, paving the way for the study described in Section \ref{Section III}, where the use of these stricter values is justified. The study consists in varying the \ac{IES} between $0.5\lambda$ and $1.5\lambda$, with $0.05\lambda$ step, resulting in the variation of $L$, varying $D$ from $40$ to $2450\lambda$. The maximum $D$ value corresponds to roughly half the shortest Fraunhofer distance, i.e. the one for the shortest considered chamber array, which has an \ac{IES} of $0.5\lambda$. The \ac{FoM} values are computed for each $L$ and $D$ combination and evaluated against the \ac{FoM} acceptable values.

The goal is to find an \ac{OTA} test setup which is as compact as possible in terms of antenna size $L$ and chamber dimensions, roughly defined by $D$. The considered values of $D$ are significantly lower than the shortest Fraunhofer distance. It is worthwhile to note that no optimization has been carried out, so the chosen combinations of $L$ and $D$ can be further improved, e.g., by the use of more advanced tapering techniques and by applying actual computational optimization techniques like the ones used in \cite{Mohammad1,Mohammad2}.

As shown in \cite{P_ANDRES_TAPER}, linear tapering is an effective technique to reduce \ac{EF} variations. Indeed, the $L$ and $D$ values fulfilling the \ac{FoM} limits were significantly lower than without tapering. The results from this study are shown in Fig.~\ref{F2}. Note that each point marked in yellow corresponds to one of the 5 chosen $L$ and $D$ combinations for Sections \ref{Section III} and \ref{Section IV}. From these results, it can be observed that, due to the highly non-linear nature of the aggregation of the \ac{EF} generated by these 100 sources, while still being in the near-field, there are discontinuities in the $L$ and $D$ combinations. I.e., for a given value of $L$, intuitively one would think that if a value of $D$ fulfills the \ac{FoM} limits, then a larger value of $D$ should fulfill them too, but that is not generally the case. On the other hand, there is some continuity in the ratios of $L$ and $D$ that fulfill the \ac{FoM} limits, forming a series of somewhat continuous ``curves" that fulfill the \ac{FoM}. Finally, if we focus only on the most compact possible setups, i.e., lowest $L$ and $D$ combinations, marked in yellow, it can be observed that there is a trade-off between $L$ and $D$, so a smaller chamber array requires a larger distance, and vice versa.

%\subsubsection{Subsubsection Heading}
%Subsubsection text here.

\section{Study of chamber array excitation errors }
\label{Section III}
\subsection{Error model}
\label{Section III A}

A study of excitation errors of the chamber array can be found, e.g., in \cite{Hybrid_exci,Exci_phased_arr}. The study conducted in this paper aims at quantifying the chamber array excitation error that can be tolerated by the selected $L$ and $D$ combinations. The error model in this paper is different from the ones presented in the references above, thus the results are not directly compared. In \cite{Hybrid_exci} two normally distributed random variables with the same standard deviation were used, which is good for simplicity. Nevertheless, the error is made proportional to the weighting coefficient, which adds complexity to the analysis. In \cite{Exci_phased_arr}, separated amplitude and phase error normally distributed random variables are assumed, each with its own standard deviation, which is impractical for the study conducted in this paper, since this would make this study two-dimensional, unnecessarily increasing the complexity.

For the sake of simplicity, the error model in this paper comprises a normally distributed complex random variable given by
\begin{equation}
    \epsilon_{ch_i}=\mathcal{N}\left(0,\sigma_{ch}\right)+j\mathcal{N}\left(0,\sigma_{ch}\right),
\end{equation}
where $\epsilon_{ch_i}$ is the excitation error of the $i$-th element of the chamber array, so a different realization of the excitation error is used for each of the \acp{AE} of the chamber array, and $\sigma_{ch}$ is the standard deviation of the excitation error. For similarity with \cite{Hybrid_exci}, the standard deviation will be increased in dB-scale $\sigma_{ch_{dB}}$. Hence,
\begin{equation}
    \sigma_{ch}={10}^{\sigma_{ch_{dB}}/20}-1.
\end{equation}
Therefore, the \ac{EF} expression at a given point $P$, is now computed as
\begin{equation}
    E_z=\sum_{i=1}^{N_C}{\left(1+\epsilon_{ch_i}\right)t_{c_i}\frac{e^{-jkr_i}}{4\pi r_i}}.
\end{equation}
\subsection{Simulation results}
\label{Section III B}

\begin{table}[b]
    % % increase table row spacing, adjust to taste
    \renewcommand{\arraystretch}{1.3}
    % if using array.sty, it might be a good idea to tweak the value of
    % \extrarowheight as needed to properly center the text within the cells
    \caption{Results of Monte-Carlo simulations of tolerable $\sigma_{ch_{dB}}$ for $L$ and $D$ combinations from Fig.~\ref{F2} (b) and Fig.~\ref{F2} (c)}
    \centering
    % % Some packages, such as MDW tools, offer better commands for making tables
    % % than the plain LaTeX2e tabular which is used here.
    \begin{tabular}{|c|c|c|c|c|}
    \hline
    $L$ & IES & $D$ & $\sigma_{ch_{dB}}$ & \ac{FoM} failed \\
    \hline
    1.43 m / 133.65$\lambda$ & 1.35$\lambda$ & 3.06 m / 286$\lambda$ & 0.05 dB & $R_{mag}$ \\
    \hline
    1.27 m / 118.8$\lambda$ & 1.2$\lambda$ & 4.73 m / 441$\lambda$ & 0.12 dB & $R_{mag}$ \\
    \hline
    1.06 m / 99$\lambda$ & $\lambda$ & 5.03 m / 469$\lambda$ & 0.11 dB & $R_{mag}$ \\
    \hline
    0.74 m / 69.3$\lambda$ & 0.7$\lambda$ & 6.04 m / 564$\lambda$ & 0.24 dB & $R_{phs}$ \\
    \hline
    0.74 m / 69.3$\lambda$ & 0.7$\lambda$ & 6.33 m / 591$\lambda$ & 0.5 dB & $R_{phs}$ \\
    \hline
    \end{tabular}
    \label{T1}
\end{table}

Monte-Carlo simulations were conducted for each selected combination of $L$ and $D$ shown in Fig.~\ref{F2} (b) and (c), where $\sigma_{ch_{dB}}$ was progressively increased with $0.01$~dB step. The \ac{FoM} were computed at each step and compared against the limits, i.e., $\sigma_{mag}\le0.25$ dB, $R_{mag}\le1$ dB, and $R_{phs}\le10^{\circ}$. When any of the \ac{FoM} does not comply with these values, then the value of $\sigma_{ch_{dB}}$ of the previous iteration is stored. The reason of not using the $L$ and $D$ from Fig.~\ref{F2} (a), is that some of the best $L$ and $D$ combinations in terms of the compactness of the setup, did not even tolerate a $\sigma_{ch_{dB}}$ value of $0.01$~dB, so more restrictive values of \ac{FoM} were used in Fig.~\ref{F2} (b) and (c) to ensure some headroom for excitation errors. The results of this study are shown in Table \ref{T1}, where each of the five points marked in Fig. \ref{F2} (b) and (c) corresponds, in order, with each of its rows. It can be observed that not all the $L$ and $D$ combinations from Fig.~\ref{F2} (b) tolerate the same amount of standard deviation of the excitation error, even though all of them, in absence of excitation error, fulfill the same level of \ac{FoM}. Additionally, the fifth $L$ and $D$ combination is, as one could intuitively think, the one with more resilience to this error, due to its larger headroom in \ac{FoM} levels.

\section{Study of \ac{DUT} weight errors}
\label{Section IV}

\subsection{Setup}
\label{Section IV A}

\begin{figure}
\centering
\includegraphics[width=0.6\columnwidth]{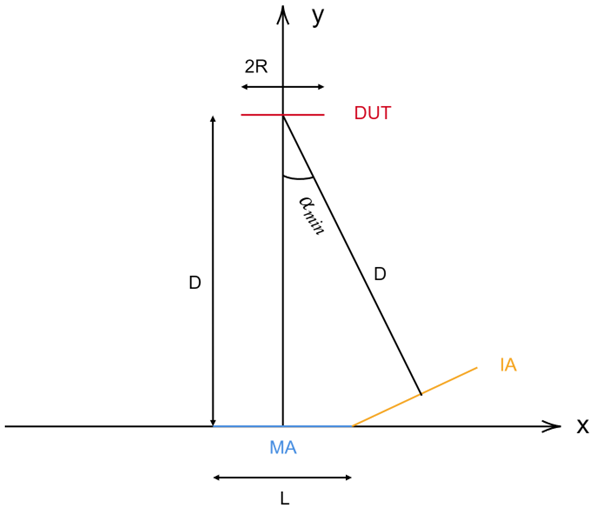}
\caption{MF and ZF performance study setup}
\label{F3}
\end{figure}

In this study, we consider a setup similar to the one shown in Fig.~\ref{F1}. Here, a second chamber array was used to emulate an interferer, and thus it is denoted as \ac{IA}. By design, it is identical to the \ac{MA} and is placed at its side, as can be seen from Fig.~\ref{F3}. The chosen $L$ and $D$ define the minimum angle $\alpha_{min}$. A generic \ac{DUT} array antenna is placed in the \ac{TZ}, and parallel to the main chamber array. The \ac{DUT} consists of $49$ vertically polarized idealized isotropic \acp{AE}. The considered $L$ and $D$ combinations are the ones from Section \ref{Section II} and are shown in Fig.~\ref{F2} and Table~\ref{T1}. Additional placements of the \ac{IA} are considered by increasing $\alpha_{min}$ by 15$^{\circ}$ for each $L$ and $D$ combinations.

\begin{figure}
    \centering
        \subfloat[]{\includegraphics[width=0.85\columnwidth]{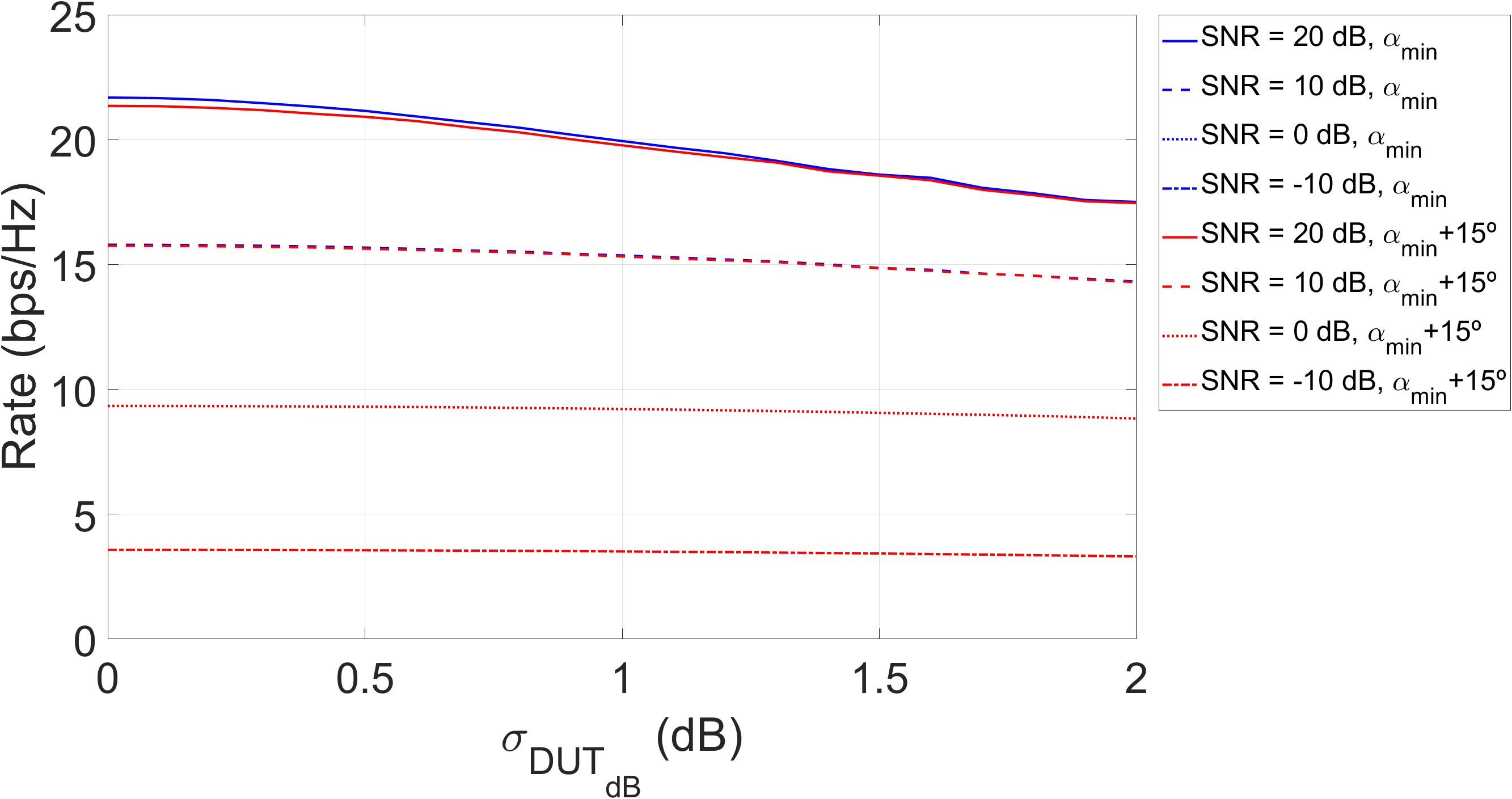}}
        \label{F4a}
        \subfloat[]{\includegraphics[width=0.85\columnwidth]{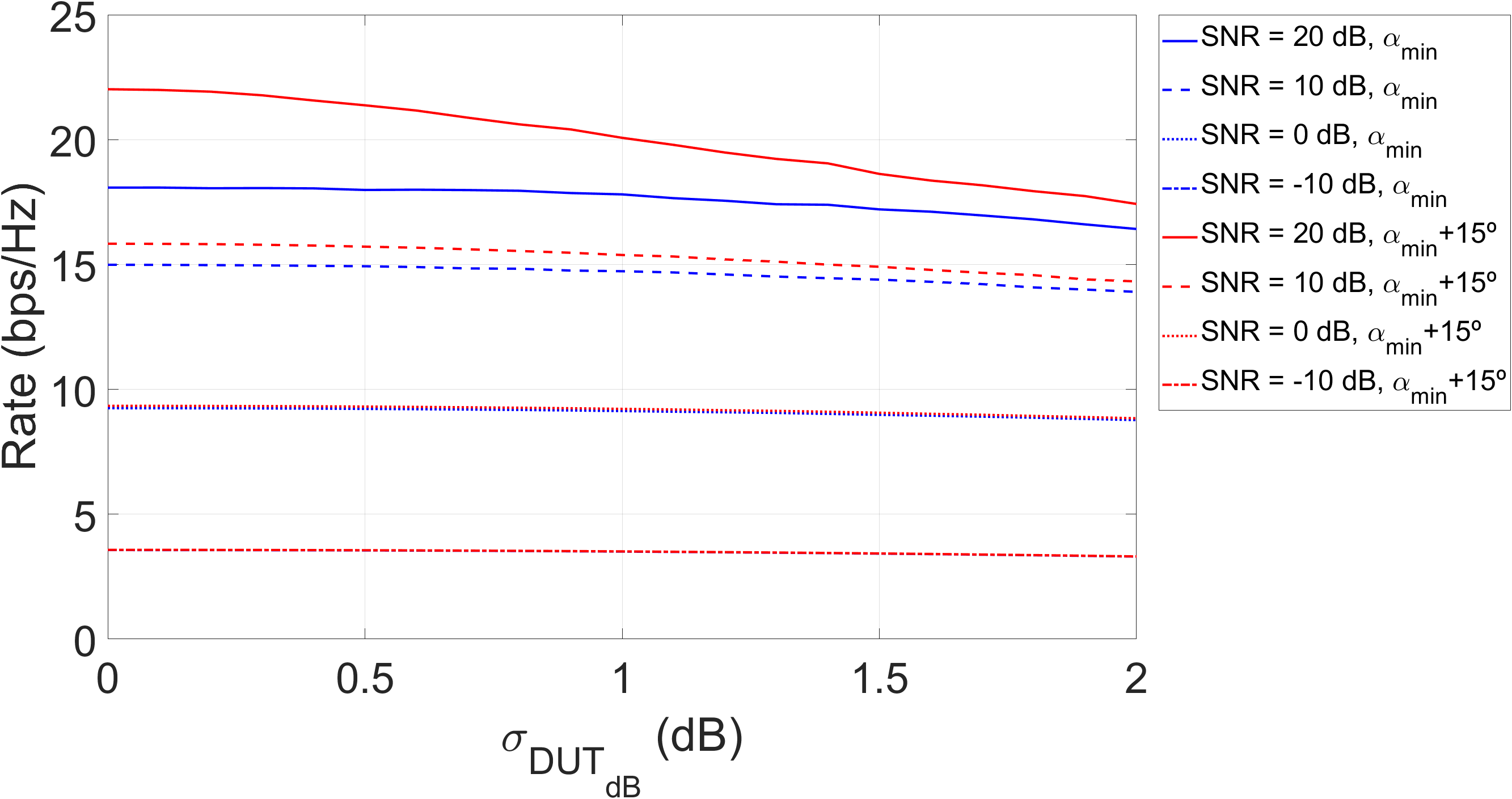}}
        \label{F4b}
    \caption{\ac{MF} average sum rate as a function of $\sigma_{DUT_{dB}}$, for different \ac{SNR} levels and for both positions of the \ac{IA} ($\alpha_{min}$ and $\alpha_{min}+15^{\circ}$). (a) $L$ and $D$ from first row of Table \ref{T1}: $L$ = $133.65\lambda$, $D$ = $286\lambda$. (b) $L$ and $D$ from fifth row of Table \ref{T1}: $L$ = $69.3\lambda$, $D$ = $591\lambda$.}
    \label{F4}
\end{figure}

\begin{figure}
    \centering
        \subfloat[]{\includegraphics[width=0.8\columnwidth]{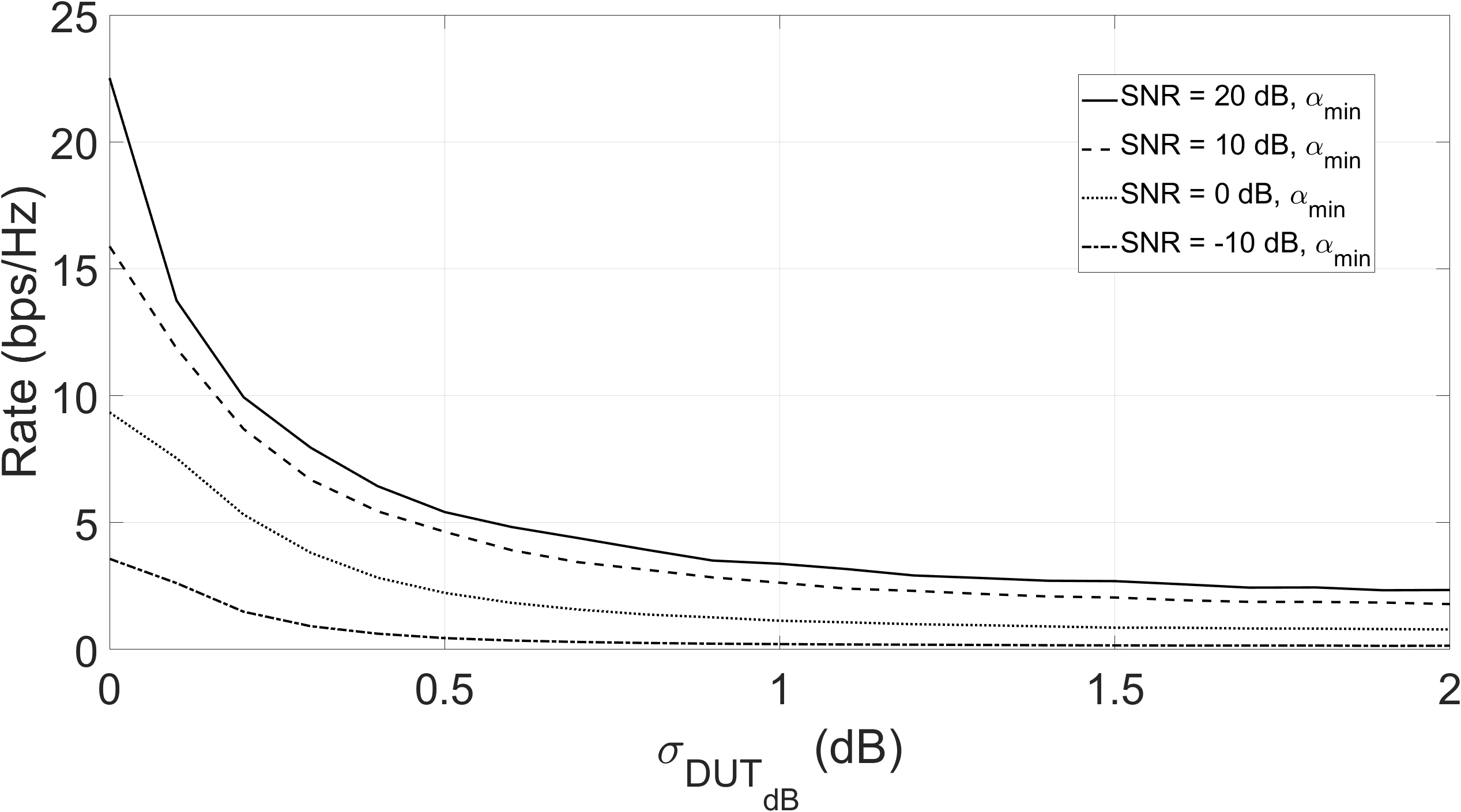}}
    \caption{\ac{ZF} average sum rate as a function of $\sigma_{DUT_{dB}}$, for different \ac{SNR} levels. $L$ and $D$ from first row of Table \ref{T1}: $L$ = $133.65\lambda$, $D$ = $286\lambda$.}
    \label{F5}
\end{figure}

\subsection{Method}
\label{Section IV B}

In this section, we aim at analyzing the impact of weight errors at the \ac{DUT} on the performance of evaluated precoding algorithms. For this purpose, we assume that the \ac{MA} and the \ac{IA} emulate two user equipment, while the \ac{DUT} is going to play the role of a base station. It could also represent two access points communicating with the onboard (on a vehicle) communications unit. Furthermore, in order to assess the impact of weight errors, we are going to measure the uplink sum rate. In order to do this, the channel matrix $\mathbf{H}$ is first computed according to (\ref{Eq1}), obtaining the $E_z$ values at each \ac{AE} of the \ac{DUT}, from the \ac{MA} and the \ac{IA}. After that, the weights are computed for both \ac{MF} and \ac{ZF}, according to:
\begin{equation}
    \mathbf{W}_{MF}=\mathbf{H}^\dagger,
\end{equation}
\begin{equation}
    \mathbf{W}_{ZF}=\mathbf{H}^\dagger(\mathbf{H}\mathbf{H}^\dagger)^{-1}.
\end{equation}

After this, $\mathbf{W}$ is distorted by an error statistically distributed as the one already presented in Section \ref{Section III A}. However, now the standard deviation is named as $\sigma_{DUT_{dB}}$, and varied from $0-2$~dB, with $0.1$~dB step. The other variable is the \ac{SNR}, which is evaluated at $-10$, $0$, $10$ and $20$ dBs. The \ac{SINR} for the \ac{MA} and the \ac{IA} is computed for each combination of $\sigma_{DUT_{dB}}$ and \ac{SNR} and, after that, the sum rate is computed according to
\begin{equation}
    SR=\sum_{u=1}^{2}\log_2{\left(1+SINR_u\right)},
\end{equation}
where the subscript $u$ refers to the chamber arrays (\ac{MA} and \ac{IA}). The same procedure was repeated for all the iterations of the Monte-Carlo simulations, with the corresponding averaging of the sum rate afterward. This is repeated for all the $L$ and $D$ combinations, as well as the different \ac{IA} placements.

\subsection{Results}
\label{Section IV C}

As expected, the performance of \ac{MF} and \ac{ZF} is impacted by the weighting errors of the \ac{DUT} array. However, it depends on the \ac{SNR}, and, as can be seen from Fig.~\ref{F4} and Fig.~\ref{F5}, this impact is much higher for \ac{ZF} than for \ac{MF}. The results for all $L$ and $D$ combinations show some differences in some cases and only for \ac{MF}, finding no relevant changes for \ac{ZF}. As for the two considered angles $\alpha_{min}$ and $\alpha_{min}+15^{\circ}$, they follow the same trend as the $L$ and $D$ combinations, having only relevant differences for some of the $L$ and $D$ combinations and only for \ac{MF}. Therefore, to illustrate the most relevant case in terms of difference of results, the results for $L$ and $D$ combinations corresponding to the first and fifth rows of Table~\ref{T1} for \ac{MF} are shown in Fig.~\ref{F4} (a) and Fig.~\ref{F4} (b), respectively, with the two considered angles. For \ac{ZF}, due to the similarity of all results, i.e. for all $L$ and $D$ combinations and both considered angles, only the $L$ and $D$ combination belonging to the first row of Table~\ref{T1} and only for the $\alpha_{min}$ angle is presented in Fig.~\ref{F5}.

In Fig.~\ref{F4}, it can be seen that the impact of the \ac{MF} weighting errors on the sum rate is only relevant for large \acp{SNR}. Additionally, the $L$ and $D$ combination of Fig.~\ref{F4} (a) shows that the performance of \ac{MF} is very similar for both positions of the \ac{IA}. On the other hand, for the $L$ and $D$ combination of Fig.~\ref{F4} (b), the performance of \ac{MF} suffers more.

In Fig.~\ref{F5}, the impact of \ac{ZF} weighting errors on sum rate is relevant for all \acp{SNR}, with similar behaviour, although affecting more to higher \acp{SNR}. In any case, the weighting error impact is larger for \ac{ZF} than for \ac{MF} across the board.

\section{Conclusion}

In this paper, we have shown that the errors due to the chamber array excitation weights may affect the feasible size of the chamber array and the distance between the test zone center to the chamber array center, i.e., the size of the testing facility. It was concluded that not all feasible combinations of these two parameters will be resilient to the chamber array errors to the same degree. It was also shown the impact on the performance of weighting coefficients errors of the \ac{DUT} array in terms of sum rate for matched filter and zero-forcing precoding algorithms, concluding that zero-forcing is, in general, much more sensitive than a matched filter to such errors. It was also concluded that the impact of weighting errors increased with the signal-to-noise ratio. We hope that the current paper paves the further steps toward future over-the-air testing solutions, especially for automotive applications. Testing solutions using FR2 frequencies for communications will benefit from the presented findings when focusing on having an over-the-air testing solution that is as compact as possible while considering different sources of error in the design process. 


\section*{Acknowledgment}
The work of Alejandro Antón was conducted within the ITN-5VC project, which is supported by the European Union’s Horizon 2020 research and innovation program under the Marie Skłodowska-Curie grant agreement No. 955629.

% \IEEEtriggeratref{7}
% \IEEEtriggercmd{\enlargethispage{-20cm}}


%\bibliographystyle{IEEEtran}

\bibliographystyle{IEEEtran}

\bibliography{References}
%\begin{thebibliography}{1}

%\bibitem{Maxwell}
%J.~Clerk~Maxwell, \emph{A Treatise on Electricity and Magnetism}, 3rd~ed., vol. 2.\hskip 1em plus 0.5em minus 0.4em\relax  Oxford: Clarendon, 1892, pp.68-73.

%\bibitem{Young}
%M.~Young, \emph{The Technical Writer's Handbook}, Mill Valley, CA: University Science, 1989.

%\bibitem{EasonNobleSneddon}
%G.~Eason, B.~Noble, and I.N.~Sneddon, ``On certain integrals of Lipschitz-Hankel type involving products of Bessel functions,'' Phil. Trans. Roy. Soc. London, vol. %A247, pp. 529-551, April 1955.

%\bibitem{Elissa}
%K.~Elissa, ``Title of paper if known,'' unpublished.

%\bibitem{Nicole}
%R.~Nicole, ``Title of paper with only first word capitalized,'' J. Name Stand. Abbrev., %in press.

%\bibitem{Yorozu}
%Y.~Yorozu, M.~Hirano, K.~Oka, and Y.~Tagawa, ``Electron spectroscopy studies on %magneto-optical media and plastic substrate interface,'' IEEE Transl. J. Magn. Japan, %vol. 2, pp. 740-741, August 1987 [Digests 9th Annual Conf. Magnetics Japan, p. 301, %1982].


%\end{thebibliography}

\end{document}